 \documentclass[prl,amsmath,amssymb,twocolumn, showpacs, superscriptaddress,10pt]{revtex4-1}

\usepackage{amsmath}
\usepackage{hyperref}
\usepackage{graphicx}
\usepackage{amsfonts}
\usepackage{amsthm}

\usepackage{color}
\definecolor{Blue}{rgb}{0.00, 0.00, 1.00}
\definecolor{Red}{rgb}{1.00, 0.00, 0.00}

\newcommand{\be}{\begin{equation}}
\newcommand{\ee}{\end{equation}}
\newcommand{\bea}{\begin{eqnarray}}
\newcommand{\eea}{\end{eqnarray}}

\newcommand{\beq}{\begin{equation}}
\newcommand{\eeq}{\end{equation}}
\newcommand{\beqn}{\begin{eqnarray}}
\newcommand{\eeqn}{\end{eqnarray}}

\begin{document}

\title{Large deviations for the height  in 1D Kardar-Parisi-Zhang growth at late times}

\author{Pierre Le Doussal}
\affiliation{CNRS-Laboratoire de Physique Th\'eorique de l'Ecole Normale Sup\'erieure, 24 rue Lhomond, 75231 Paris Cedex, France}
\author{Satya N. \surname{Majumdar}}
\affiliation{LPTMS, CNRS, Univ. Paris-Sud, Universit\'e Paris-Saclay, 91405 Orsay, France}
\author{Gr\'egory \surname{Schehr}}
\affiliation{LPTMS, CNRS, Univ. Paris-Sud, Universit\'e Paris-Saclay, 91405 Orsay, France}

\date{\today}

\begin{abstract}
We study the atypically large deviations of the height $H \sim {{\cal O}}(t)$ at the origin at late times in $1+1$-dimensional
growth models belonging to the Kardar-Parisi-Zhang (KPZ) universality class. We present exact results
for the rate functions for the discrete single step growth model, as well as for the continuum KPZ equation
in a droplet geometry. Based on our exact calculation of the rate functions we argue that
models in the KPZ class undergo a third order phase transition from a strong coupling to
a weak coupling phase, at late times. 
\end{abstract}
\maketitle





The celebrated Tracy-Widom (TW) distribution was discovered originally in random matrix theory (RMT) \cite{TW94,TW96}. 
In RMT, it describes the probability distribution of the typical fluctuations of the largest eigenvalue of
a Gaussian random matrix. Since then, this distribution has emerged in a variety of problems~\cite{TW_review,Wolchover} 
(unrelated a priori to RMT), ranging
from random permutations \cite{baik} all the way up to the Yang-Mills gauge field theory~\cite{FMS11}. 
Why is TW distribution so ubiquitous? It was recently shown that in several systems where TW distribution occurs there is usually an underlying {\it third order} phase transition between a strong 
and a weak coupling phase \cite{rmt_review}. In these systems the TW distribution appears as a finite-size crossover function connecting the free-energies
of the two phases across the third order critical point~\cite{rmt_review}. In the strong coupling phase, the degrees of freedom of the system act collectively
while the weak coupling phase is described by a single dominant degree of freedom. In the context of RMT, this third-order phase transition 
shows up in the distribution of the top eigenvalue $\lambda_{\max}$ of a $N \times N$ matrix 
belonging to the classical Gaussian ensembles~\cite{rmt_review}. 
The central part of the distribution, corresponding to the {\it typical} fluctuations of $\lambda_{\max}$, is described by the TW distribution, 
while the {\it atypically} large fluctuations to the left (right) correspond to the strong (respectively weak) 
coupling phases. 

For a wide class of 1+1-dimensional interface growth models belonging to 
the Kardar-Parisi-Zhang (KPZ) universality class \cite{KPZ}, it is well known that 
the {\it typical} height fluctuations grow at late times as $\sim t^{1/3}$
\cite{directedpoly}. 
Moreover the probability distribution function (PDF) of these typical fluctuations
are given by the TW distribution \cite{johansson,spohn2000,TW2001,Majumdar2004,MN2005,SS10,CLR10,DOT10,ACQ11,reviewCorwin}. 
This TW distribution has also been verified experimentally in liquid crystal and paper burning
systems 
\cite{takeuchi,myllys,HH-TakeuchiReview}. The appearance of the TW distribution in these growth models
then raises a natural question: is there a third order phase transition between a strong and a 
weak coupling phase in such growth models? If so, how can one describe these two phases? In this Letter, we 
show that there is indeed a third order phase transition in these growth models by studying the probability distribution $P(H,t)$
of the height $H$ at the origin (suitably centered) 
at late times $t \gg 1$. Specifically, we find that $P(H,t)$, for $t \gg 1$, has
three different behaviors  
\begin{eqnarray}\label{main_results}
\hspace*{-1cm}P(H,t) \sim
\begin{cases}
& e^{-t^2\,\Phi_-(H/t)} \quad , \quad  H \sim {\cal O}(t)  < 0 \; \quad {\rm I} \label{1} \\
&\\
&\dfrac{1}{t^{1/3}} f\left[ \dfrac{H}{t^{1/3}} \right] \;, \; \hspace*{0.3cm} 
H \sim {\cal O}(t^{1/3}) \quad {\rm II} \label{2} \\
&\\
& e^{-t \,\Phi_+(H/t)} \quad , \quad  H \sim {\cal O}(t) > 0 \quad {\rm III} \label{3} \;.
\end{cases}
\end{eqnarray}
\begin{figure}  
\begin{center}
\includegraphics[width = \linewidth]{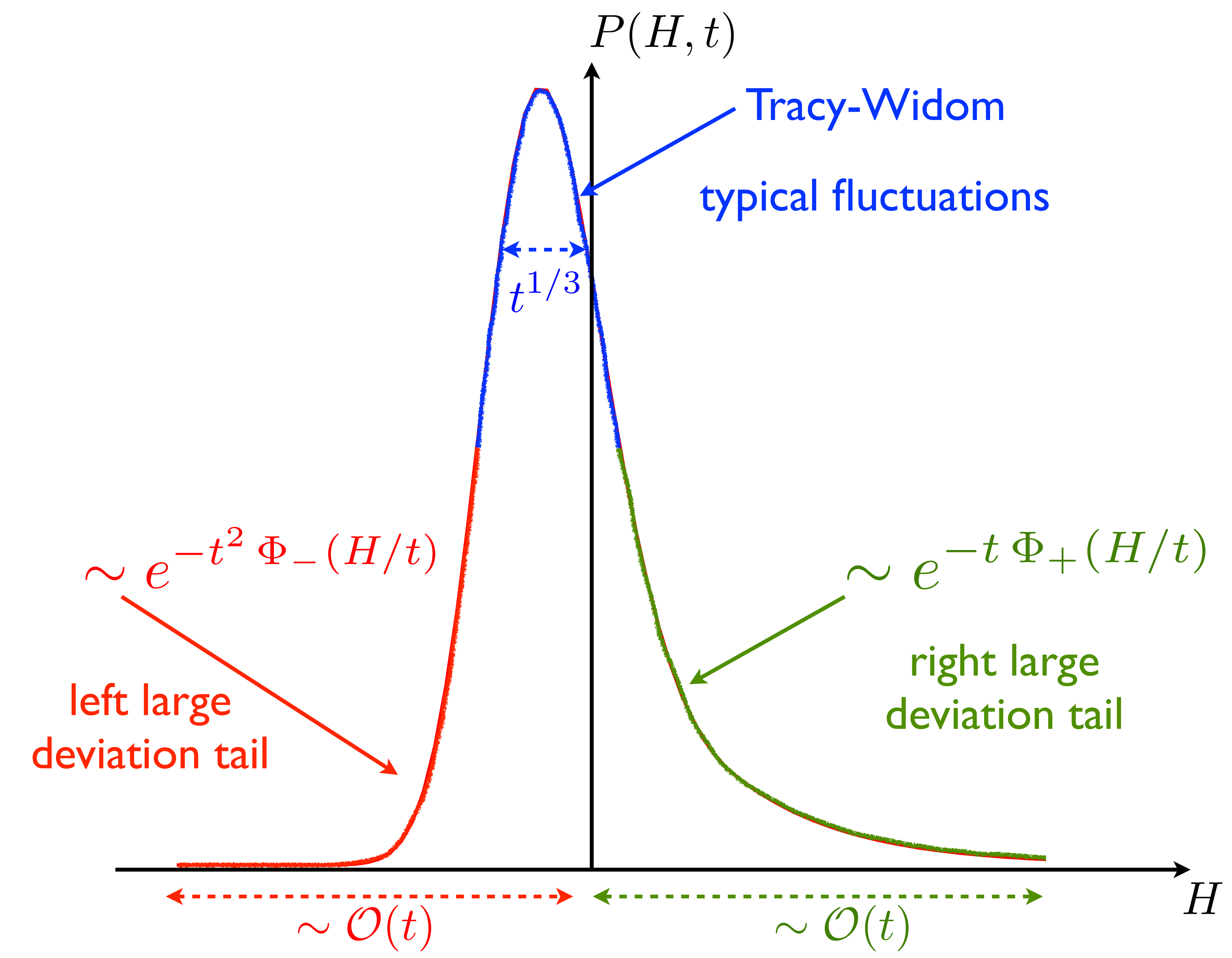}
\caption{A schematic picture of the height distribution at the
origin. The typical fluctuations $H \sim {\cal O}(t^{1/3})$ around the mean are 
distributed according to the Tracy-Widom GUE 
law (blue line). The atypical large fluctuations to the left (red line)
and to the right (green line) are described respectively by the
left and right large deviation functions in Eq.~(\ref{main_results}).}
\label{fig:height_dist}
\end{center}
\end{figure}
The regime II is well known and it describes the typical height fluctuations ($H \sim {\cal O}(t^{1/3})$) and the scaling function $f(s)$ is given by the TW
distribution. The scaling function depends on the initial conditions: for the flat geometry it corresponds to $f_1(s)$ ({\it i.e.}, TW for the Gaussian Orthogonal Ensemble, GOE), while for the curved (or droplet) geometry, it corresponds to $f_2(s)$ ({\it i.e.}, TW for the Gaussian Unitary Ensemble, GUE). These distributions have asymmetric non-Gaussian tails:
\begin{eqnarray}\label{asympt_TW}
f_\beta(s) \sim
\begin{cases}
&e^{-\frac{\beta}{24} |s|^3} \;, \; s \to - \infty \\
&e^{-\frac{2\beta}{3} s^{3/2}} \;, \; s \to + \infty \;,
\end{cases}
\end{eqnarray}
where $\beta = 1$ and $2$ correspond respectively to GOE and GUE. 

The new results in this Letter concern the atypical large height fluctuations in 
regime I and III in Eq. (\ref{main_results}). The regime I corresponds to the 
large negative fluctuations ($H \sim {\cal O}(t) < 0$) and is characterized by the 
left large deviation function $\Phi_-(z)$. Similarly, the regime III describes the large 
positive fluctuations ($H \sim {\cal O}(t) > 0$) and is 
characterized by the right large deviation function $\Phi_+(z)$. These two rate 
functions $\Phi_\pm(z)$ are the characteristics of the two phases: $\Phi_-(z)$ corresponds to the 
strong coupling phase, while $\Phi_+(z)$ describes the weak coupling phase 
(as explained later). Note that on the scale 
$H \sim {\cal O}(t)$, the central part of width ${\cal O}(t^{1/3})$ 
is effectively reduced to a point $z=0$ as $t \to \infty$.
Indeed, it follows from Eq. (\ref{main_results}) that 
\begin{eqnarray} 
\lim_{t \to \infty} -\frac{1}{t^2} \ln P(H = z \, t,t) =
\begin{cases}
&\Phi_-(z)\;, \;  z \leq 0\\
&0 \;, \; z \geq 0 \;.
\end{cases}
\end{eqnarray}
{\it Thus as $t\to \infty$, $z=0$ becomes a critical point} and $\Phi_-(z)$ can be interpreted as the 
``free energy'' of the strong coupling phase. We further show that it vanishes universally, 
$\Phi_-(z) \propto |z|^3$, as $z \to 0^-$, thus indicating a third order phase transition. 
Therefore in order to probe this third order transition it is important to compute the large deviation 
functions. In this Letter, we compute $\Phi_{\pm}(z)$ explicitly for the droplet geometry 
in (i) a discrete single step growth model belonging to the KPZ class and (ii) the continuum KPZ equation. 
In general, $\Phi_{\pm}(z)$ are non-universal and depend on the model. However, their small 
arguments behaviors are universal: $\Phi_-(z) \propto |z|^3$ as $z \to 0^-$ and $\Phi_+(z) \propto 
z^{3/2}$ as $z \to 0^+$. Indeed, as the critical point $z=0$ is approached from either side, 
the large deviation behaviors smoothly match with the asymptotic tails of the TW distribution 
(\ref{asympt_TW}).        

\begin{figure}
\begin{center}
\includegraphics[width = \linewidth]{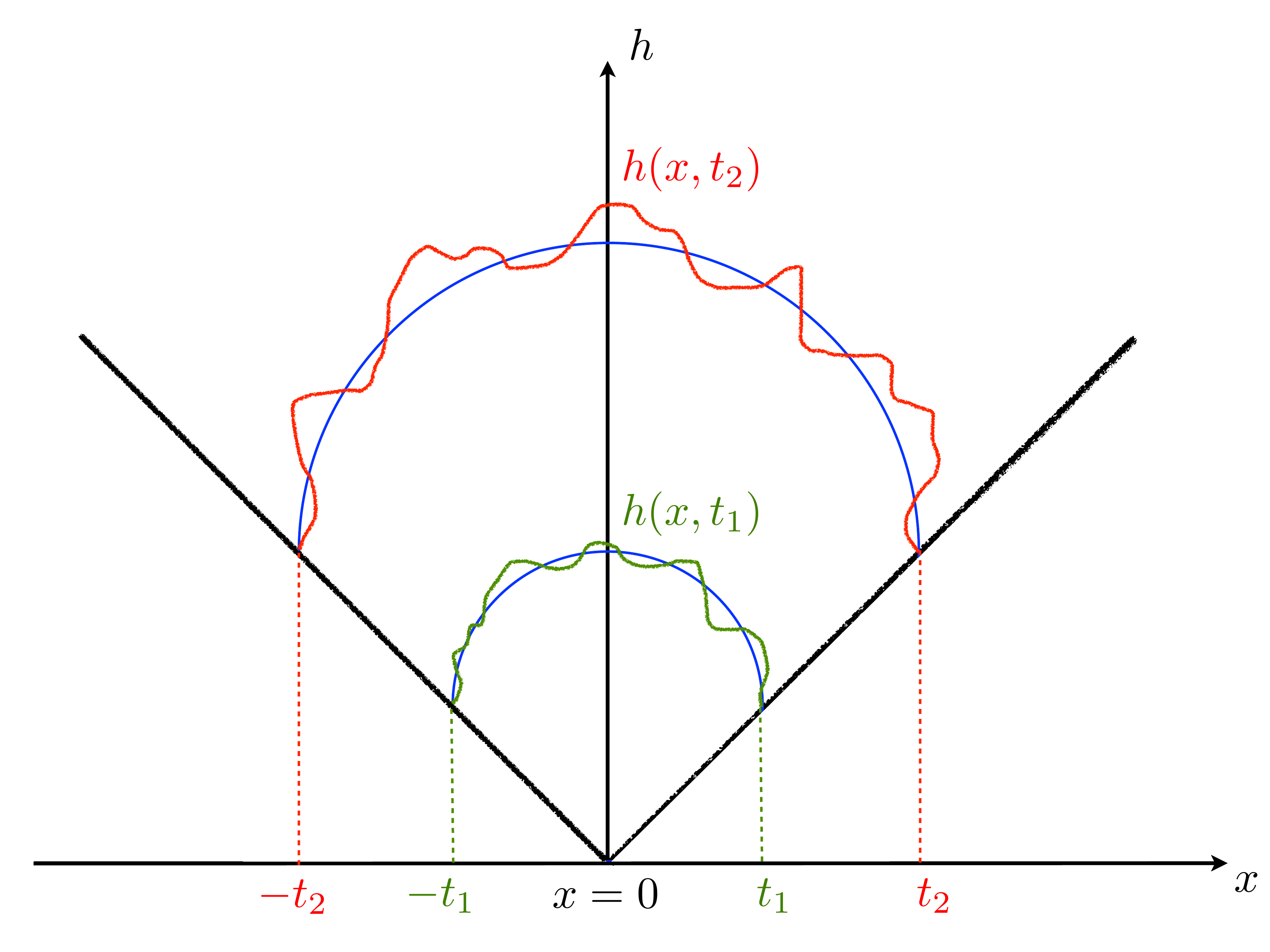}
\caption{The height $h(x,t)$ evolving on a substrate $-t\le 
x\le t$. The light cone (black lines) describes the evolution 
of the substrate. The solid line (in blue) represents the
average height at two different times, $\langle h(x,t)\rangle= v(x/t)\, t$ with
$v(z)= 1+\sqrt{1-z^2}$ having a semi-circular shape.}
\label{fig:height1:text}
\end{center}
\end{figure}
We start by analyzing a directed polymer model belonging to the KPZ universality class 
studied by Johansson \cite{johansson}. This model can be translated to a discrete
space-time $(x,t)$ growth model in a ``droplet" geometry. The
growth takes place on the substrate $-t\le x\le t$ (see 
Fig.~\ref{fig:height1:text}), starting from the seed at the origin $x=0$ at $t=0$. 
The interface height $h(x,t)$, at site $x$ and at time $t$, evolves in the bulk $-t<x<t$ as~\cite{footnote_edge}
\begin{equation}
h(x,t) = {\rm max}[ h(x-1,t-1), h(x+1,t-1)] + \eta(x,t)\,\label{bulk_evol_text} 
\end{equation}
where $\eta(x,t)\ge 0$'s are independent and identically distributed (i.i.d.) 
nonnegative random 
variables each drawn from an exponential distribution: $p(\eta)=e^{-\eta}$
for $\eta\ge 0$. Johansson showed that at late times, the average height
$\langle h(x,t)\rangle= v(x/t)\, t$ with
$v(z)= 1+\sqrt{1-z^2}$ exhibiting a semi-circular droplet shape (see Fig. \ref{fig:height1:text}).
Moreover the height at the origin at late times behaves as $h(0,t) \simeq 2 \,t + 2\,t^{1/3} \chi_2$,
where $\chi_2$ is a $t$-independent random variable distributed via the TW distribution for the GUE, $f_2(s)$ \cite{johansson}. By exploiting
an exact mapping to the largest eigenvalue of complex Wishart matrices \cite{johansson}, and using the results
for the large deviations of the latter \cite{VMB,MV}, we establish the result in Eq. (\ref{main_results}) (with $H=\frac{h(0,t)}{2}- t$). In regime I, we get \cite{supp_mat}:    
\begin{eqnarray}
\Phi_-(z) = \frac{1}{8}\left(2z-z^2 - 2\,\ln{(1+z)} \right) \;,\; -1< z  \leq 0 \;, \label{phi-_joh} 
\end{eqnarray}
where $z>-1$ since the height $h(0,t) > 0$. As $z \to 0^-$, one gets $\Phi_-(z) \sim |z|^3/12$ as announced in the introduction. 
In regime III, we find
\begin{equation}
\Phi_+(z) = 2 \sqrt{z(z+1)} + \ln{\left( 2z + 1 - 2 \sqrt{z(z+1)}\right)}  \;, \; z \geq 0 \label{phi+_joh} \;,
\end{equation}
which behaves as $\Phi_+(z) \sim (4/3) z^{3/2}$ as $z \to 0^+$. Note that in regime II, if we make $H \sim {\cal O}(t)$ and use the asymptotic behaviors
of TW distribution in Eq. (\ref{asympt_TW}) with $\beta = 2$, it can be checked that it matches smoothly with the large deviation regimes on both sides. Interestingly, in this height model (\ref{bulk_evol_text})  there is a clear physical explanation as to why the left tail [regime I in (\ref{main_results})] scales like $\sim e^{-t^2}$ while the right tail [regime III in (\ref{main_results})] behaves like $\sim e^{-t}$. Indeed, in order to realize a configuration of $H$ much smaller than its typical value (regime I), the noise variables $\eta(x,t)$ at {\it all the sites} within the $1+1$-dimensional wedge (cf Fig. \ref{fig:height1:text}) should be small. Indeed, if any of the $\eta(x,t)$ within this wedge is big, the dynamics in Eq. (\ref{bulk_evol_text}) would force the neighboring sites at the next time step to be big. The probability of this event, where {\it collectively} all the noise variables $\eta(x,t)$ inside the wedge ($|x|<t$), of area $\propto t^2$, are all small is proportional to $e^{-t^2}$ (the noise variables being i.i.d.). In contrast, a configuration where $H$ is much bigger than its typical value (regime III) can be realized by adding large positive noise variables at the origin $\eta(x=0,\tau)$ at all times $\tau$ between $0$ and $t$. The probability of this event is simply $\propto e^{-t}$ as the noises at different times are i.i.d. Hence this event is not a collective one, unlike the left large deviation. Thus, the left large deviation [regime I in Eq. (\ref{main_results})] is the analogue of the `strong coupling phase' and the right 
large deviation [regime II in Eq. (\ref{main_results})] corresponds to the `weak coupling' 
phase. The 
transition between the two phases is a third order phase 
transition, as $\Phi_-(z) \propto |z|^3$ as $z \to 0^-$, as mentioned above. This picture is very similar
to other third order phase transitions observed before in RMT and reviewed
recently in Ref.~\cite{rmt_review}.

While the right tail rate function $\Phi_+(z)$ has been studied
numerically \cite{Monthus} and, more recently, analytically
\cite{LogGamma} in discrete growth models, the left tail
$\Phi_-(z)$ is much harder to compute, and there are
very few exact results, an exception being 
the longest increasing subsequence in random permutations (for both tails) \cite{Zeitouni}. 
We now show that these rate functions can also be calculated
for the continuum KPZ equation itself, 
where the height field $h(x,t)$ evolves as \cite{KPZ}
\be
\label{eq:KPZ}
\partial_t h = \nu \, \partial_x^2 h + \frac{\lambda_0}{2}\, (\partial_x h)^2 + \sqrt{D} \, \xi(x,t) \;,
\ee
where $\nu > 0$ is the coefficient of diffusive relaxation, $\lambda_0 > 0 $ is the strength of the 
non-linearity and $\xi(x,t)$ is a Gaussian white noise with zero mean and 
$\langle \xi(x,t) \xi(x',t')\rangle = \delta(x-x')\delta(t-t')$. We use everywhere
the natural units of space $x^*=(2 \nu)^3/(D \lambda_0^2)$, time $t^*=2(2 \nu)^5/(D^2 \lambda_0^4)$ and height
$h^*=\frac{2 \nu}{\lambda_0}$.

Here for definiteness we focus on the narrow wedge initial condition, 
$h(x,0) = - |x|/\delta$, with $\delta \ll 1$,
which gives rise to a curved (or {\em droplet}) 
mean profile as time evolves
~\cite{SS10,CLR10,DOT10,ACQ11,reviewCorwin}.
We focus on the shifted height at the origin at $x=0$, $H(t) = h(0,t) + \frac{t}{12}$,
which fluctuates {\it typically} 
on a scale $t^{1/3}$ around its mean at large time, as described by the regime II in Eq.~(\ref{main_results}) with $f(s) = f_2(s)$, the TW 
distribution for the GUE. 
We show below that for the continuum KPZ equation, in a droplet geometry, the generic result in Eq. (\ref{main_results}) holds in regime I and III
as well. Interestingly the rate functions turn out to be rather simple in this case
%
\begin{eqnarray}\label{eq:left_tail1}
&& \Phi_-(z) = \frac{1}{12} |z|^3  \quad , \quad z \leq 0 \\
&& \Phi_+(z) = \dfrac{4}{3} z^{3/2}  \quad , \quad z \geq 0 \;. \label{eq:left_tail2}
\end{eqnarray}
Thus the continuum KPZ equation also exhibits a third order phase transition at
the critical point $z=0$.

To derive the rate functions for the continuum KPZ case, 
we start from an exact formula \cite{SS10,CLR10,DOT10,ACQ11}, valid at all times $t$ in the
droplet geometry. It relates the following generating function to a Fredholm determinant (FD) 
%
%
\be  
g_{t}(s) :=  \langle \exp(- e^{H(t) - t^{1/3} s })\rangle \; = \det(I - P_s K_{t} P_s) \label{GF}
\ee
where the finite time kernel is 
\bea
&&K_{t}(r,r') = \int_{-\infty}^\infty du \frac{{\rm Ai}(r+u) {\rm Ai}(r'+u)}{1+ e^{- t^{1/3} u}} 
 \label{eq:Airy_def} \;
\end{eqnarray}
and $P_s$ is the projector on the interval $[s,+\infty)$ \cite{footnote_Fredholm}. In Eq. (\ref{eq:Airy_def}), ${\rm Ai}(x)$
denotes the Airy function.

Let us recall that to obtain the typical fluctuations regime
(II) in formula (\ref{2}), where $H(t) \sim t^{1/3}$, one needs to 
 take the
limit $t \to +\infty$ at fixed $s$ in (\ref{GF}). In that limit $K_t(r,r')$ converges to the standard Airy kernel, 
$K_{\rm Ai}(r,r') = \int_{0}^\infty du \, {\rm Ai}(r+u) {\rm Ai}(r'+u)$ and the right hand side (r.h.s.) converges
to the GUE-TW distribution. The left hand side (l.h.s.)~of (\ref{GF}) converges to 
$\langle \theta(s- t^{-1/3} H(y))\rangle$ (where $\theta(x)$ is the Heaviside step function),
and one obtains
\be 
\lim_{t \to + \infty}\, {\rm Prob.} ( \chi_t < s ) = \det(I - P_s K_{Ai} P_s) = F_2(s)
\ee 
where $F_2(s)=\int_{-\infty}^s f_2(s') ds'$ is the cumulative distribution function (CDF)
of the GUE-TW distribution. To compute the rate functions $\Phi_{\pm}(z)$ we now consider the 
formula (\ref{GF}) in the limit when $s$ and $t$ are both large, 
keeping the ratio $y=s/t^{2/3}$ fixed. 

{\it Right tail.} 
We start with the right large deviation function, therefore we consider formula (\ref{GF}) 
in the regime of large $s>0$. Consider first the l.h.s. of Eq. (\ref{GF}). 
It is convenient to introduce a random variable $\gamma$ (independent of $H$) 
distributed via the Gumbel distribution, of
CDF given by
\bea \label{gumbel}
\langle \theta(b-\gamma) \rangle_\gamma = e^{-e^{-b}} \;.
\eea 
Substituting $b=s t^{1/3}-H$ in (\ref{gumbel}) allows us to rewrite
the l.h.s of (\ref{GF}) as
\be \label{leftgf} 
 1-\langle \exp(- e^{H(t) - t^{1/3} s })\rangle
 =  \langle {\rm Prob}(H > s t^{1/3} - \gamma) \rangle_\gamma \;.
\ee
Now consider the r.h.s of Eq. (\ref{GF}) 
for $s \gg 1$. Expanding the FD in powers of $K_t$
and keeping only the first two terms one obtains
\be \label{trace} 
\det(I - P_s K_{t} P_s)  \simeq 1 - \int_{s}^{+\infty} dr K_t(r,r) \;.
\ee
Equating Eq. (\ref{leftgf}) and (\ref{trace}) and taking a derivative
with respect to $s$ gives 
\bea \label{gamma} 
t^{1/3} \langle P(H = s t^{1/3} - \gamma,t) \rangle_\gamma = K_t(s,s) 
\eea 
a relation exact for all $t$.

We first study the asymptotics of $K_t(s,s)$ for large $s \sim t^{2/3}$. 
Performing a change of variable $u = - t^{2/3} v$, (\ref{eq:Airy_def}) becomes
\bea \label{kernel_asymp} 
K_{t}(y t^{2/3}, y t^{2/3}) = t^{2/3} \int_{-\infty}^{+\infty} dv \frac{{\rm Ai}^2(t^{2/3}(y-v))}{1+e^{t v}}
\eea 
with $y={\cal O}(1)$. This integral can be analyzed for large $t$ \cite{supp_mat}
and we obtain \cite{footnote2}
%
\begin{equation}
K_t(y t^{2/3}, y t^{2/3}) \sim e^{- t I(y)} \;, \; I(y) =  \begin{cases}
\frac{4}{3} y^{3/2} \, , \, 0<y<\frac{1}{4} \\
y - \frac{1}{12} \, , \, y> \frac{1}{4} \;,
\end{cases} \label{F0} 
\end{equation}
where the pre-exponential factors are given in \cite{supp_mat}. Having obtained
the r.h.s of (\ref{gamma}) we now consider its l.h.s. We anticipate (and verify a posteriori)
that in this right tail the PDF has the form (setting $z=H/t$)
%
\be\label{subdom0} 
\ln P(H,t) = - t \frac{4}{3} z^{3/2} - a \ln t  - \chi_{\rm droplet}(z) + o(1) 
\ee
where the constant $a$ and the function $\chi_{\rm droplet}(z)$ are yet to be determined. 
Inserting this form on the l.h.s. of Eq.~(\ref{gamma}), analyzing the resulting
integral \cite{supp_mat} and comparing it to the r.h.s. in (\ref{F0}),
we find that indeed the ansatz in (\ref{subdom0}) is correct 
with $a=1$ and an explicit form for $\chi_{\rm droplet}(z)$ given 
in Eq. (\ref{chi_droplet_less}) of the Supp. Mat.~\cite{supp_mat}. Finally,
keeping only the leading behavior of (\ref{subdom0})
gives us the exact right rate function
%
\bea \label{legendre3} 
\Phi_+(z) = \frac{4}{3} z^{3/2} \quad , \quad z \geq 0 \;,
\eea 
as announced in Eq. (\ref{eq:left_tail2}). For the pre-exponential
factor in the flat case we find $a=1/2$ and $\chi_{\rm flat}(z)$ given 
in Eq.~(\ref{flat_3}) of the Supp. Mat.~\cite{supp_mat}.

This result is also consistent with the known exact large time 
behavior of the moments, 
$\overline{e^{ n H} } \sim_{t \to +\infty} e^{\frac{1}{12} n^3 t}$, 
calculated using the Bethe ansatz \cite{kardareplica}. Indeed a saddle point calculation using 
$P(H,t) \sim e^{- \frac{4}{3} (\frac{H}{t^{1/3}})^{3/2}}$ reads
\bea
\int dH ~ e^{ n H - \frac{4}{3} (\frac{H}{t^{1/3}})^{3/2} } \sim e^{\frac{1}{12} n^3 t}
\eea 
where the saddle point, at $H_n = n^2 t/4$ for fixed integer $n$, is
precisely in the right large deviation regime. Note that the dependence on the 
initial condition appears only in the (subdominant) pre-exponential factor
of the moments, as discussed in \cite{supp_mat} where we establish that 
$\Phi_+(z)= \frac{4}{3} z^{3/2}$ both for droplet and flat initial conditions. 

{\it Left tail.} We now focus on the left tail where we set $H/t \sim {\cal O}(1) <0$. 
In this case, one can show \cite{supp_mat} that the l.h.s. of (\ref{GF}) scales as
$\sim e^{- t^2 \Phi_-(y=s/t^{2/3})}$ for $y={\cal O}(1)$. The r.h.s of 
(\ref{GF}), $Q_t(s) := \det(I - P_s K_{t} P_s)$, 
is not easy to analyze in the regime of large negative $s$.
Fortunately in Ref. \cite{ACQ11} the authors proved an exact differential equation
satisfied by $Q_t(s)$: 
\bea \label{painleve1} 
\partial_s^2 \ln Q_t(s) = - \int_{-\infty}^{+\infty} dv \sigma'_t(v) [q_t(s,v)]^2
\eea 
where 
\bea
\sigma_t(v) = \frac{1}{1 + e^{- t^{1/3} v}} 
\eea 
and $\sigma_t'(v)=\partial_v \sigma_t(v)$. The function $q_t(s,v)$ satisfies a non-linear integro-differential equation
in the $s$ variable 
\be  \label{painleve2} 
\partial_s^2 q_t(s,v) = (s+v+ 2  \int_{-\infty}^{+\infty} dw \sigma'_t(w) [q_t(s,w)]^2 ) q_t(s,v)
\ee
with the boundary condition $q_t(s,v) \simeq_{s \to +\infty} {\rm Ai}(s+v)$. 
In the long limit $t \to +\infty$, $\sigma_t'(v) \to \delta(v)$
and hence $q_t(s,0)$ satisfies the standard Painlev\'e II equation
\cite{TW94}.

For large but finite $t$, we substitute the anticipated scaling form $Q_t(s) \sim e^{-t^2 \Phi_-(y=s/t^{2/3})}$
in (\ref{painleve1}). The consistency then suggests that $q_t(s,v)$ takes the scaling form
\bea \label{scalingq} 
q_t(s,v) \simeq t^{1/3} \tilde q(s/t^{2/3},v t^{1/3}) \,, \; {\rm for} \; t \; \to \infty \;,
\eea 
and the scaling function $\tilde q(y,v)$ satisfies 
\bea \label{36} 
\int_{-\infty}^{+\infty} dv \frac{\tilde q(y,v)^2 e^{-v}}{(1 + e^{- v})^2} = \Phi''_-(y).
\eea 
Substituting further the scaling form (\ref{scalingq}) in the
differential equation (\ref{painleve2}) we obtain as $t \to \infty$
\bea
y + 2 \int_{-\infty}^{+\infty} dv \frac{\tilde q(y,v)^2 e^{-v}}{(1 + e^{- v})^2} = 0 \;.
\eea 
Comparing with (\ref{36}) immediately gives for all $z \leq 0$, $\Phi''_-(z)  = - \frac{z}{2}$. 
Solving with the boundary condition $\Phi_-(z) \simeq_{z \to 0} |z|^3/12$, 
coming from matching with the left tail of the TW GUE distribution as 
$z \to 0^-$, implies
\bea 
\Phi_-(z) = \frac{1}{12} |z|^3  \quad , \quad z \leq 0 \;,
\eea 
as announced in Eq. (\ref{eq:left_tail1}).

In summary, our results on large deviations for the height at late times for
growth models in the KPZ class suggest a third order phase transition 
between a strong and a weak coupling phase. Generically
the associated rate functions are non-universal but their small
argument behavior are universal, as they match the TW tails. 
In the case of the continuum KPZ equation these functions
are simple, Eqs. (\ref{eq:left_tail1}, \ref{eq:left_tail2}), showing that the TW universality extends 
all the way to the large deviation regime. A natural question is how 
this late time behavior is approached as time increases. 
Weak noise expansion and instanton calculations in the tails (for the flat geometry) 
indicate a different behavior $P(H,t) \sim e^{- |H|^{5/2}/t^{1/2}}$ in the left tail in the early time 
regime $t \ll1$ \cite{Baruch,Korshunov}. In fact we have computed 
exactly the short time height distribution in the droplet geometry
which exhibits a similar $|H|^{5/2}$ left tail behavior \cite{us},
manifestly different from the late time behavior $|H|^{3}$ obtained here at late times. In contrast,
the right tail $H^{3/2}$ is already attained at early time.

We thank D. Dean, 
B. Meerson, J. Quastel, H. Spohn and
K. Takeuchi
for useful discussions. We acknowledge support from PSL grant ANR-10-IDEX-0001-02-PSL
(PLD). 
We thank the hospitality of KITP, under Grant No. NSF PHY11-25915.


{}

\newpage

.

\begin{widetext} 

\bigskip

\bigskip

\begin{large}
\begin{center}

SUPPLEMENTARY MATERIAL

\end{center}
\end{large}

\bigskip

\section{Johansson model} 

Johansson's directed polymer model in $2$-dimensions~\cite{johansson} is defined as follows.
Consider a $2$-d lattice where a site $(i,j)$ has a quenched energy $\eta(i,j)$, drawn independently
for each site from an exponential distribution: $p(\eta)= e^{-\eta}$ with $\eta\ge 0$. Consider now
a directed path from the origin to the site $(M,N)$ ($M\ge 0$, $N\ge 0$). The energy of a path is just the sum
of the energies of all sites belonging to the path. From all possible paths ending at $(M,N)$, one
considers the optimal path, {\it i.e.}, the one with the highest energy. Let $E(M,N)$ denote the
energy of this optimal path. One can easily write a recursion relation
\begin{equation}
E(M,N)= {\max}[E(M-1,N), E(M,N-1)] + \eta(M,N) \,.
\label{energy_recur.1}
\end{equation}
Clearly, $E(M,N)$ is a random variable and one is interested in its probability distribution.
Making the change of variables, $x=M-N$ and $t=M+N$ and denoting $E(M,N)\equiv h(x,t)$, it reduces
to an interface growth model, where the height $h(x,t)$ ($-t\le x\le t$), evolves with discrete time $t$ according to
the following rules (see Fig. \ref{fig:height1:text}),
\begin{equation}
h(x,t) = {\rm max}[ h(x-1,t-1), h(x+1,t-1)] + \eta(x,t)\, \quad\quad {\rm
for}\quad -t<x<t \,.
\label{bulk_evol.supp}
\end{equation}
At the two edge points $x=\pm t$, the evolution of the height $h(\pm t,t) \equiv 
h_{\pm}(t)$ is slightly different
\begin{eqnarray}
h_{+}(t) &=& h_{+}(t-1) + \eta(t,t) \label{right_evol_supp} \\
h_{-}(t) &=& h_{-}(t-1) + \eta(-t,t) \label{left_evol_supp} \;.
\end{eqnarray}
At late times, the average height at point $x$ converges to~\cite{johansson}
\begin{equation}
\langle h(x,t)\rangle \to v\left(\frac{x}{t}\right)\, t\,; \quad\quad
-t\le x\le t
\label{avg_height_supp}
\end{equation}
where $v(z=x/t)= 1+ \sqrt{1-z^2}$ has a semi-circular form (see Fig. \ref{fig:height1:text}).
The height $h(x,t)$ fluctuates around this average {\em typically} on a scale $\sim {\cal O}(t^{1/3})$ for large $t$.
In particular, at $x=0$, the height at late times converges to $h(0,t) \to 2 t + 2\, t^{1/3}\, \chi_2$,
where the random variable $\chi_2$ is of ${\cal O}(1)$ (independent of $t$ for large
$t$) and is distributed via the Tracy-Widom GUE law~\cite{johansson}.
In other words, the PDF of the scaled (and centered) height at the origin
\begin{equation}
H= \frac{h(0,t)}{2}-t \, 
\label{scaled+height}
\end{equation}
has the late time
scaling form
\begin{equation}
P(H,t) \sim \frac{1}{\,t^{1/3}}\, f_2\left(
\frac{H}{t^{1/3}}\right)
\label{dist_height0}
\end{equation}
where $f_2(s)$ is the TW GUE PDF with asymptotics given in Eq. (\ref{asympt_TW}) of the main text with $\beta=2$.
This is represented schematically by the central blue region in Fig.
\ref{fig:height_dist} of the main text.

In contrast to the typical fluctuations, the atypically large fluctuations
both to the left and to the right of the mean, are not described by the
Tracy-Widom distribution. To compute these tails, one can use an exact mapping
due to Johansson~\cite{johansson} that states
\begin{equation}
{\rm Prob. \,}[E(M,N) \le l]= {\rm Prob. \,}[\lambda_{\max}\le l] \;,
\label{mapping_wishart.1}
\end{equation}
where $\lambda_{\max}$ denotes the largest eigenvalue of an $(M\times N)$ complex Wishart matrix
defined as follows. Let $X$ be an $(M\times N)$ rectangular matrix whose entries are independent
complex Gaussian variables, ${\rm Prob. \,}[X]\propto \exp[- {\rm Tr}(X^{\dagger} X)]$. Construct
then the product matrix $W= X^\dagger X$ which is $(N\times N)$ and has $N$ non-negative real
eigenvalues ${\lambda_1,\lambda_2,\ldots,\lambda_N}$ with maximal eigenvalue $\lambda_{\max}={\max}(\lambda_1,
\lambda_2,\ldots,\lambda_N)$. Without any loss of generality, one can assume $N\le M$.
The statistics of $\lambda_{\max}$ has been studied extensively
in the random matrix literature and one can then borrow these results for our problem.
In terms of the height, the relation (\ref{mapping_wishart.1}) simply reads
\begin{equation}
{\rm Prob. \,}[h(x,t) \le l]= {\rm Prob. \,}[\lambda_{\max}\le l] \;,
\label{mapping_wishart.2}
\end{equation}
where $x= M-N$ and $t=M+N$. Since we are interested in the height at $x=0$, this corresponds to
the Wishart matrix with $M=N$ and $N=t/2$. For $M=N$, it is well known~\cite{johansson} that for large $N$,
$\lambda_{\rm max}\to 4\, N + 2^{4/3}\, N^{1/3}\, \chi_2$. Using $N=t/2$, one immediately recovers
the result that $h(0,t) \to 2\, t+ 2\, t^{1/3}\, \chi_2$ for large $t$ as mentioned above.
In addition, the large deviation tails of $\lambda_{\max}$ for Wishart matrices are also
known~\cite{VMB,MV}. For $M=N$, they read as $N\to \infty$
\begin{eqnarray}
{\rm Prob. \,}[\lambda_{\max}\le l] &\sim & \exp\left[- N^2\, \psi_{-}^W\left(\frac{l}{N}\right)\right]
\quad\quad\quad\quad {\rm for}\quad 0\le \frac{l}{N}\le 4 \label{left_wishart.1}  \\
& \sim & 1- \exp\left[-N\, \psi_{+}^W\left(\frac{l}{N}\right)\right]
\quad\quad \;\;\, {\rm for}\quad \frac{l}{N}\ge 4 \label{right_wishart.1}
\end{eqnarray}
where the left rate function $\psi_{-}^W(y)$ is given explicitly as~\cite{VMB}
\begin{equation}
\psi_{-}^W(y)= \ln 4- \ln y - \left(1-\frac{y}{4}\right) -\frac{1}{2} \left(1-\frac{y}{4}\right)^2 \qquad\qquad\qquad\quad\quad\, {\rm for}\quad 0\le y\le 4 \;,
\label{left_wishart_vmb}
\end{equation}
while the right rate function $\psi_{+}^W(y)$ has the expression~\cite{MV}
\begin{equation}
\psi_{+}^W(y)= -\ln 4 + \sqrt{y(y-4)} + 2\,\ln\left(y-2-\sqrt{y(y-4}\right) \quad\quad {\rm for}\quad y\ge 4 \;.
\label{right_wishart_mv}
\end{equation}
Note that the superscript $W$ stands for Wishart matrices. 

To translate these results to the height model and derive the large deviation results mentioned
in Eq. (\ref{main_results}) in the main text, we consider the scaled height
defined in Eq. (\ref{scaled+height}). Then, using $N=t/2$, we get
\begin{equation}
{\rm Prob. \,}[H\le t\, z]= {\rm Prob. \,}[h(0,t)\le 2\,(1+z)\,t]= {\rm Prob. \,}[\lambda_{\max}\le 4\,(1+z)\,N] \;.
\label{translate.1}
\end{equation}
Finally, using the results from Eqs. (\ref{left_wishart.1}) and (\ref{right_wishart.1}) and using
again $N=t/2$ we obtain the announced results
\begin{eqnarray}
{\rm Prob. \,}[H\le t\, z] &\sim & \exp\left[- t^2\, \Phi_{-}(z)\right]
\quad\quad\quad {\rm for}\quad -1\le z \le 0 \label{left_ldv1.supp}  \\
& \sim & 1- \exp\left[-t\, \Phi_{+}(z)\right]
\quad\quad {\rm for}\quad z\ge 0 \label{right_ldv1.supp}
\end{eqnarray}
where the rate functions $\Phi_{\pm}(z)$ can be expressed explicitly in terms of the Wishart rate functions
in Eqs. (\ref{left_wishart_vmb}) and (\ref{right_wishart_mv}). We get
\begin{eqnarray}
\Phi_{-}(z)& = & \frac{1}{4}\psi_{-}^W(4\,(1+z))= \frac{1}{4}\,\left[z-\frac{z^2}{2}-\ln(1+z)\right] \quad \hspace*{2.72cm} {\rm for}
\quad -1\le z\le 0 \;, \label{left_ldv2_supp} \\
\Phi_{+}(z) & =& \frac{1}{2}\,\psi_{+}^W (4\,(1+z))= 2\,\sqrt{z(1+z)} + \ln\left(2z+1-2\sqrt{z(1+z)}\right) \,
\quad {\rm for}
\quad z\ge 0  \;. \label{right_ldv2_supp}
\end{eqnarray}
Taking derivatives with respect to $z$ in Eqs. (\ref{left_ldv1.supp}) and (\ref{right_ldv1.supp}), one gets the
large deviation tails of the PDF $P(H,t)$ of the scaled height $H$ at the origin as announced 
in Eqs. (\ref{phi-_joh}) and (\ref{phi+_joh}) respectively in the main text. 

Note that while the large deviation principle in this problem was originally established by Johansson
\cite{johansson}, the left rate function $\Phi_-(z)$
was not computed. Here we obtain this function explicitly
in (\ref{left_ldv2_supp}). While a general expression for the right rate function $\Phi_+(z)$ was 
computed by Johansson for the geometric disorder, here we obtain a simplified explicit expression
for $\Phi_+(z)$ in (\ref{right_ldv2_supp}) for the exponential disorder.

\vskip 0.3cm

\noindent{\bf {Matching with the tails of the Tracy-Widom distribution:}}

\vskip 0.2cm

We start from the left tail.
When the scaled height $z=H/t$ in Eq. (\ref{scaled+height}) approaches $0$ from below, it is easy to see by expanding
$\Phi_{-}(z)$ to leading order for small $z$
\begin{equation}
\Phi_{-}(z) \sim \frac{|z|^3}{12}\, .
\label{left_matching1.supp}
\end{equation}
Substituting this result in Eq. (\ref{left_ldv1.supp}) and taking a derivative with respect to $z$, one finds that when $z\to 0^-$, the 
left large deviation tail of the PDF of $H$ behaves as
\begin{equation}
P(H,t)\sim \exp\left[-\frac{|H|^3}{12\,t}\right]\,. 
\label{left_matching2.supp}
\end{equation}
On the other hand, if we start from the central Tracy-Widom distribution that
describes typical fluctuations of ${\cal O}(t^{1/3})$ in Eq. 
(\ref{dist_height0}), and set $H=z\,t$,
we will probe the probability of fluctuations to the left that are much 
larger (of ${\cal O}(t)$) than the typical size ${\cal O}(t^{1/3})$. This gives
\begin{equation}
P(H=z\,t, t) \sim t^{-1/3}\, f_2\left(z\, t^{2/3}\right)  \;.
\label{ltw.1}
\end{equation}
As $t\to \infty$ with fixed $z<0$, the argument of $f_2$ in Eq. (\ref{ltw.1})
tends to negative infinity. So, we need to use the left tail asymptotic
of the Tracy-Widom density in Eq. (\ref{asympt_TW}) of the main text:
$f_2(s)\sim \exp[- |s|^3/12]$. Substituting this in Eq. (\ref{ltw.1})
gives $P(H, t)\sim \exp[-|H|^3/12 t]$, which matches smoothly
with the result in Eq. (\ref{left_matching2.supp}) obtained from
the small argument behavior of the left large deviation regime.

A similar matching can be verified on the right side as well. When $z$ approaches $0^+$ from above,
we get by expanding $\Phi_+(z)$ to leading order
\begin{equation}
\Phi_+(z) \sim \frac{4}{3}\, z^{3/2}\, . 
\label{right_matching1.supp}
\end{equation}
Substituting in Eq. (\ref{right_ldv1.supp}) and taking a derivative with respect to $z$, one finds that when $z\to 0^+$, the
right large deviation tail of the PDF of $H$ behaves as
\begin{equation}
P(H,t)\sim \exp\left[-\frac{4}{3}\, \frac{H^{3/2}}{\sqrt{t}}\right]\,.
\label{right_matching2.supp}
\end{equation}
In contrast, starting from the central TW regime (valid on a scale $H\sim t^{1/3}$), and setting $H=z\,t$ gives
Eq. (\ref{ltw.1}) where $z>0$.
As $t\to \infty$ with fixed $z>0$, the argument of $f_2$ in Eq. (\ref{ltw.1})
now tends to positive infinity. Hence, we use the right tail asymptotic
of the Tracy-Widom density in Eq. (\ref{asympt_TW}) of the main text:
$f_2(s)\sim \exp[- \frac{4}{3}\, s^{3/2}]$. Substituting this in Eq. (\ref{ltw.1})
gives $P(H, t)\sim \exp[-\frac{4}{3}\,\frac{H^{3/2}}{\sqrt{t}}]$, which then matches smoothly
with the result in Eq. (\ref{right_matching2.supp}) obtained from
the small argument behavior of the right large deviation regime.

Note that although here we have restricted ourselves, for simplicity, to the height at the origin $x=0$, the computations presented above 
can be easily extended to the large deviations of the height $h(x,t)$ at a generic point~$x$. 

\section{Right tail asymptotics of the kernel at equal points} 
\label{sec:rightkernel} 

In Eq. (\ref{kernel_asymp}) of the main text, for the simplicity of reading, 
we only provided the leading exponential factor
for the asymptotic expansion of the kernel. However, one can easily obtain also
the subdominant pre-exponential factors as shown below. 

We start by evaluating the asymptotic behavior of the integral on the
r.h.s. of (\ref{kernel_asymp}) with $y={\cal O}(1)$ fixed and as $t\to \infty$. 
It turns out that the dominant contribution to this integral
comes from the interval $v \in [-\infty,y]$. In this interval, for large $t$, we
can replace the Airy function by its large positive tail
asymptotics ${\rm Ai}(z) \simeq \frac{1}{\sqrt{4 \pi z^{1/2}}} e^{- \frac{2}{3} z^{3/2}}$
as $z \to +\infty$. This leads to 
\bea \label{kernel2} 
&& K_{t}(y t^{2/3}, y t^{2/3}) \simeq \frac{t^{1/3}}{4 \pi} \int_{-\infty}^y \frac{dv}{(y-v)^{1/2}} 
\frac{ e^{- t \frac{4}{3} (y-v)^{3/2} } }{1 + e^{v t}} \;.
\eea
It turns out that there are two regimes (i) $y>1/4$ (ii) $0< y< 1/4$. 

In the first regime $y>1/4$, the integral can be evaluated by the saddle point method.
We first assume, and then check a posteriori, that there is a saddle point $v^*>0$. Then
the integral will be dominated near $v^*>0$. Then,
one can replace $1/(1+ e^{v t})$ by $e^{-v t}$ for large $t$ with $v>0$ and
evaluate the integral by the saddle point method: 
\bea
&& K_{t}(y t^{2/3}, y t^{2/3}) \simeq \frac{t^{1/3}}{4 \pi} \int_{-\infty}^y \frac{dv}{(y-v)^{1/2}} e^{- t S(y,v)} \\
&& S(y,v) = \frac{4}{3} (y-v)^{3/2} + v  \;. \label{legendre1} 
\eea 
For $y>1/4$, the saddle point is at $v^*= y - \frac{1}{4}$. For consistency we need $v^*>0$, i.e., $y>1/4$.
Evaluating the integral at this saddle point gives
\bea \label{Kr} 
&& K_t(y t^{2/3}, y t^{2/3}) \simeq 
\frac{1}{\sqrt{4 \pi t^{1/3}}} e^{- t (y- \frac{1}{12})} \;.
\eea 

In the second regime $0<y<1/4$, there is no saddle point and the dominant contribution to the
integral in (\ref{kernel2}) comes from the edge $v \approx 0$. 
Setting $v=w/t$ and keeping only leading order terms for large $t$ we obtain
\bea
 K_{t}(y t^{2/3}, y t^{2/3}) \simeq \frac{e^{- t \frac{4}{3} y^{3/2} } }{4 \pi t^{2/3} \sqrt{y}} 
 \int_{-\infty}^{+\infty} dw \frac{e^{2 \sqrt{y} w}}{1 + e^w} 
\eea 
This integral can be performed explicitly giving
\bea \label{Kr2} 
K_{t}(y t^{2/3}, \, y t^{2/3}) \simeq \frac{e^{- t \frac{4}{3} y^{3/2} } }{4  t^{2/3} \sqrt{y}\, 
\sin\left(2\pi\sqrt{y}\right)}\,. 
\eea

If we neglect the pre-exponential factors we recover the
formula given in the text, namely
\bea
&& K_t(y t^{2/3}, y t^{2/3}) \sim 
e^{- t I(y)} \\
&& I(y) =  \begin{cases}
\frac{4}{3} y^{3/2} \quad , \quad 0<y<\frac{1}{4} \\
y - \frac{1}{12} \quad , \quad y> \frac{1}{4} \;.
\end{cases} \label{F} 
\eea

\section{Pre-exponential factor in the right large deviation tail}

Inspired by the form of the subdominant corrections in
the right large deviation tail of the top eigenvalue of a Gaussian
random matrix \cite{subdominant}, it is natural to make the following ansatz
in the limit of large time
\bea \label{subdom} 
\ln P(H,t) \simeq { - t \frac{4}{3} z^{3/2} - a \ln t  - \chi(z) + o(1) } \quad , \quad z=H/t \quad \text{fixed} \;.
\eea 
In this section we establish this behavior, both using moments from the replica
method and using the exact form of the generating function. We also
calculate $a$ and $\chi(z)$ explicitly, both for the flat as well as droplet initial conditions
and show that they do depend on the initial conditions.

\subsection{Moments from the replica method} 

The positive integer moments of $e^H$ for the continuum KPZ equation
can be studied using the mapping to the attractive Lieb-Liniger model with $n$ bosons \cite{kardareplica}. 
 From the Bethe ansatz solution of this model the exact formula for the moments 
 at arbitrary time \cite{CLR10,DOT10}
takes the form of a sum of exponentials
\bea \label{moment1} 
\langle e^{ n H} \rangle = \sum_{\mu_n} B_{\mu_n,t} ~ e^{ - E_{\mu_n} t } 
\eea
where the index $\mu_n$ labels the $n$- boson eigenstates. In the limit of
large system size $L=+\infty$, these are made of so-called strings, with a total energy spectrum 
\bea
E_{\mu_n} = \sum_{j=1}^{n_s} m_j k_j^2 - \frac{1}{12}  m_j^3 \quad , \quad \sum_{j=1}^{n_s} m_j=1 \quad , \quad 
m_j \geq 1 \quad , \quad
1 \leq n_s \leq n 
\eea 
where the $k_j$ are the (real) momenta of each string. 

At large time and fixed positive integer $n$, the sum (\ref{moment1}) is
dominated by the ground state $|0\rangle_{n,k=0}$, together with
its center of mass finite momentum excitation, i.e. more precisely, taking into
account the gap with the next set of excited states
\bea \label{mom1} 
\langle e^{ n H} \rangle \sim_{t \to +\infty} A_{n,t} e^{\frac{1}{12} n^3 t} [1+ {\cal O}(e^{- \frac{1}{4} n (n-1) t}) ]
\eea
where the amplitude (see e.g. \cite{PLDflat}) 
\bea
A_{n,t} = \lim_{L \to +\infty} \frac{1}{L} \sum_{k=\frac{2 \pi p}{n L}, p \in \mathbb{Z}} e^{-n k^2 t} 
\frac{1}{n^2} \langle \Psi_0 | 0 \rangle_{n,k} \;.
\eea 
The last factor is the overlap, i.e., the scalar product of the (unnormalized) ground state wave function
(such that $\langle 0,..0|0 \rangle_n = n!$), with the (unnormalized) replica wave function
$| \Psi_0 \rangle$ encoding for the initial condition. This overlap is complicated in
general, but is known for some special initial conditions. This leads for $n \geq 1$ to
\bea \label{flat1}
 \langle \Psi_0 | 0 \rangle_{n,k} = n^2 2^{n-1} L~  \delta_{k,0} \quad &\Rightarrow& \quad A_{n,t}= 2^{n-1} 
 \quad , \quad \quad\quad \hspace*{0.65cm}
\text{flat initial condition} \;, \\ \label{drop1}
 \langle \Psi_0 | 0 \rangle_{n,k} = n! \quad &\Rightarrow& \quad A_{n,t}= \frac{n!}{n^{3/2} (4 \pi t)^{1/2}} 
 \quad , \quad 
\text{ droplet initial condition} \;.
\eea 
The saddle point method described in the text can be extended to
obtain the pre-exponential factor. Substituting the anticipated form
(\ref{subdom}) we obtain for any fixed integer $n>0$ and large $t$
\bea
&& \langle e^{n H} \rangle \simeq t \int dz e^{ - t (\frac{4}{3} z^{3/2} - n z) - a \ln t - \chi(z) }  
\simeq t^{\frac{1}{2}-a} \sqrt{\pi  n} ~ e^{\frac{t n^3}{12} - \chi(n^2/4)} \;, \label{saddle1} 
\eea 
obtained using the saddle point at $z=z_n = n^2/4$. 

In the flat initial condition case, comparing (\ref{mom1}), (\ref{flat1}) with
(\ref{saddle1}) one finds $a=1/2$ and the correction to scaling function
\bea\label{flat_chi_integer}
\chi_{{\rm flat}}(z) = \frac{1}{2} \ln(8 \pi) + \frac{1}{4} \ln z -  (\ln 4) \sqrt{z}    \quad , \quad z=z_n=n^2/4 \quad , \quad n \in \mathbb{N}^* \;.
\eea
In the droplet case we get 
$a=1$ and $\chi_{\rm droplet}(n^2/4)=\ln(2 \pi n^2/n!)$, hence the correction to scaling function
\bea \label{chiinteger} 
\chi_{{\rm droplet}}(z) = \ln( 4 \pi)  + \frac{1}{2} \ln z  - \ln\left(\Gamma(2 \sqrt{z})\right) \quad , \quad z=z_n=n^2/4 \quad , \quad n \in \mathbb{N}^* \;.
\eea
%

\medskip

\subsection{From moments to the generating function} 

Expanding the generating function in Eq. (\ref{GF}) in terms of moments, reads
\bea
 g_t(s) = 1 + \sum_{n \geq 1} \frac{(-1)^n}{n!} e^{- n t^{1/3} s} \langle e^{n H} \rangle 
&\simeq& 1 - \int_{-\infty}^\infty du \, {\rm Ai}(2 u + 2^{2/3} s) (1 - e^{-2 e^{2^{1/3} t^{1/3} u}}) \quad , \quad 
\text{flat initial condition} \label{flat_int} \\
& \simeq & 1 -  \int_{s}^\infty dr \int_{-\infty}^\infty du \, \frac{{\rm Ai}(r+u)^2}{1+ e^{- t^{1/3} u}} 
 \quad , \quad \hspace*{1.1cm}
\text{droplet initial condition}\; . \label{droplet_int}
\eea 
To obtain the first line we used (\ref{mom1}), (\ref{flat1}) 
and the  "Airy trick" identity $\int_{-\infty}^{+\infty} dy {\rm Ai}(y) e^{y w} = e^{w^3/3}$ for
$w>0$. To obtain the second line we used  (\ref{mom1}), (\ref{drop1}) and the
following variant
\bea
e^{- n t^{1/3} s} 
\frac{e^{n^3 t/12}}{n^{3/2} (4 \pi t)^{1/2}} = \int_{s}^{+\infty} dr \int_{-\infty}^\infty du \, {\rm Ai}(r+u)^2 e^{n t^{1/3} u}
\eea 
for $n>0$, and then summed up the geometric series in $n$ (see \cite{CLR10} and
Section 4.2.1 in \cite{PLDflat} for details). In the droplet case it recovers the expansion
(\ref{trace}) and for $t \to +\infty$ in the flat case it also reproduces (\ref{trace}) where $K_t(r,r')$ is replaced
by the GOE kernel ${\rm Ai}(r+r')$. The asymptotics of these kernels
then allow to recover the asymptotics obtained by the saddle point
method, showing that, to obtain the right tail large deviations, it is equivalent to work on
the replica formula or on the generating function, as mentioned in the text
and also done below. 

\subsection{Right tail from the generating function: droplet initial condition} 

Taking a derivative of $g_t(s)$ with respect to $s$ in (\ref{GF}) we obtain the
relation (valid for all $t$ and large $s$)
\bea \label{exact.0} 
\langle e^{ H - s t^{1/3} - e^{H - s t^{1/3}}} \rangle = \frac{1}{t^{1/3}} K_t(s,s) \;.
\eea 
Setting $s=y t^{2/3}$ and $H=t z$ it can be rewritten as
\bea \label{exact} 
\langle e^{ t (z-y)  - e^{t(z-y)}} \rangle = \frac{1}{t^{1/3}} K_t(y t^{2/3},y t^{2/3}) \;.
\eea 
The r.h.s. of this equation has been analyzed in a Section above.
We now analyze the l.h.s. of Eq. (\ref{exact}). 

Consider first the case $y>1/4$. In the large $t$ limit, using (\ref{Kr}), the r.h.s. reads:
\bea
\frac{1}{\sqrt{4 \pi t}} e^{- t (y- \frac{1}{12})} \;.
\eea 
Inserting now the the anticipated form 
(\ref{subdom}) in the l.h.s. one sees that for $y>1/4$ it can be evaluated
by the saddle point method, the saddle point being at $z=1/4$. One 
obtains 
\bea
\langle e^{ H - s t^{1/3} - e^{H - s t^{1/3}}} \rangle 
\simeq \sqrt{\pi} t^{\frac{1}{2}-a} e^{- \chi_{\rm droplet}(1/4) - t (y- \frac{1}{12})} \;.
\eea 
Comparing the two sides we obtain $a=1$ and $\chi_{\rm droplet}(1/4)=\ln(2 \pi)$ in
perfect agreement with the replica calculation (for $n=1$). 

Let us now consider the case $0<y<1/4$. Using (\ref{Kr2}), the r.h.s. of (\ref{exact}) reads for large time
\bea \label{rhs2} 
\frac{e^{- t \frac{4}{3}\, y^{3/2} }}{4\, t\, \sqrt{y}\,\sin\left(2\pi\sqrt{y}\right) }\, .    
\eea 
Inserting now the the anticipated form 
(\ref{subdom}) in the l.h.s. of (\ref{exact}) we see that
for $0<y<1/4$ the integral is dominated by the region
of $z$ near $y$. Let us write $z=y+w/t$ and expand the integrand in powers of $t$. This gives
\bea
&& \langle e^{ H - s t^{1/3} - e^{H - s t^{1/3}}} \rangle 
\simeq \frac{1}{t^a}  e^{- t \frac{4}{3} y^{3/2} - \chi_{\rm droplet}(y) } 
\int_{-t y}^{+\infty} dw ~ e^{(1-2 \sqrt{y}) w - e^w - \frac{w^2}{2 t \sqrt{y}} }\, .
\eea
If $y<1/4$ and is kept fixed, as $t \to +\infty$, the last integral can be calculated by
neglecting the quadratic term in the exponential and setting the lower integration limit to
$-\infty$. It then becomes $\Gamma(1- 2 \sqrt{y})$. Matching now with the r.h.s
(\ref{rhs2}) gives
$a=1$ and
\bea
e^{-\chi_{\rm droplet}}\, \Gamma(1-2\sqrt{z})= \frac{1}{4\,\sqrt{z}\,\sin\left(2\pi \sqrt{z}\right)}\, .
\label{chidrop.1}
\eea
Using $\Gamma(x)\Gamma(1-x)= \pi/\sin(\pi x)$, this immediately gives
\bea
\label{chi_droplet_less}
\chi_{\rm droplet}(z) = \ln (4 \pi) + \frac{1}{2} \ln z -\ln\left(\Gamma\left(2\sqrt{z}\right)\right)\, . 
\eea 
One checks from Eq. (\ref{chi_droplet_less}) that $\chi_{\rm droplet}(z)\to \ln(2\pi)$ as $z\to 1/4$ from
below, thus matching perfectly with the result obtained for $z = 1/4$
given by Eq. (\ref{chiinteger}) for $n=1$. In fact, this formula
for $\chi_{\rm droplet}(z)$ in Eq. (\ref{chi_droplet_less}) is valid for all $z>0$, and
clearly coincides with Eq. (\ref{chiinteger}) for $z=n^2/4$ (obtained for integer $n$).

\subsection{Right tail from the generating function: flat initial condition} 



We start with the following relation, obtained from Eq. (\ref{flat_int}), 
\bea
\langle (\exp( - e^{ H - t^{1/3} s} ) - 1) \rangle = - \int_{-\infty}^{+\infty} du\, {\rm Ai}(2 u + 2^{2/3} s)\, 
(1- e^{-2 e^{(2 t)^{1/3} u}}) \;.
\eea 

Denoting $H=t z$ and making the change of variable $u = - 2^{-1/3} v t^{2/3}$, 
$s=y t^{2/3}$ we obtain
\bea
\langle (\exp( - e^{ t ( z - y)  } ) - 1) \rangle = - 2^{-1/3} t^{2/3} 
\int_{-\infty}^{+\infty} dv {\rm Ai}\big((2 t)^{2/3} (y-v) \big) 
(1- e^{-2 e^{- v t}}) \;.
\label{flat.1}
\eea 
We first evaluate the asymptotics of the r.h.s. of Eq. (\ref{flat.1}).
On the r.h.s. the dominant contribution to the integral comes from the interval
$v \in (-\infty,\, y]$. Replacing the Airy function by its large positive tail
asymptotics ${\rm Ai}(z) \simeq \frac{1}{\sqrt{4 \pi z^{1/2}}} e^{- \frac{2}{3} z^{3/2}}$
as $z \to +\infty$, we find
\bea
 - \frac{t^{1/2}}{\sqrt{8 \pi}} \int_{-\infty}^{y} dv 
 \frac{1}{ (y-v)^{1/4}}
 e^{- \frac{4}{3} t  (y-v)^{3/2} }
(1- e^{-2 e^{- v t}}) \;.
\eea 
For $0<y<1/4$, this integral is dominated by the neighborhood of $v=0$. Setting $v=w/t$, expanding
and keeping only the leading terms gives
\bea
- \frac{e^{-t\, \frac{4}{3}\, y^{3/2}}}{\sqrt{8\pi t}\, y^{1/4}}\, f_R(y), \quad\quad {\rm where}\quad 
f_R(y)= \int_{-\infty}^{\infty} dw\, e^{2\sqrt{y}\,w}\, \left[1- e^{-2\, e^{-w}}\right]\, .
\label{rhs_flat.2}
\eea

We now turn to the l.h.s of Eq. (\ref{flat.1}).
We substitute the anticipated form (\ref{subdom}) for $P(H,t)$ (with $H=zt)$ on the l.h.s of (\ref{flat.1}).
This results in the following integral
\bea
t^{1-a}\,  \int_0^{\infty} dz\, \left[e^{-e^{t(z-y)}}-1\right]\, e^{-\frac{4}{3}\,t\, z^{3/2}} \, e^{-\chi_{\rm flat}(z)}\,.
\label{lhs_flat.1}
\eea
For large $t$, this integral is dominated by the neighborhood of $z=y$. Hence, we set $z=y-w'/t$, expand
in $t$ and keep only up to leading order terms for large $t$. This gives the l.h.s 
\bea
-t^{-a}\, e^{-\chi_{\rm flat}(y)}\, e^{-\frac{4}{3}\,t\, y^{3/2}}\, f_L(y), \quad\quad {\rm where}\quad
f_L(y)= \int_{-\infty}^{\infty} dw'\, e^{2\sqrt{y}\,w'}\,\left[1- e^{-e^{-w'}}\right]\,.
\label{lhs_flat.2}
\eea
In fact, with a change of variable, it is easy to show that $f_L(y)= 2^{-\sqrt{4y}}\, f_R(y)$.

Comparing the l.h.s in (\ref{lhs_flat.2}) with the r.h.s in (\ref{rhs_flat.2}) gives
$a=1/2$ and 
\bea
\chi_{\rm flat}(z)= \ln\left(\sqrt{8\pi}\, z^{1/4}\, \frac{f_L(z)}{f_R(z)}\right)= \frac{1}{2}\ln (8\pi)+\frac{1}{4}\ln(z)-
(\ln 4)\sqrt{z}\, .
\label{flat_3}
\eea
This result is valid for all $z>0$ and matches perfectly with the result in Eq. (\ref{flat_chi_integer}) obtained from
the integer moments.

\subsection{Matching with the right tail of Tracy-Widom distributions}

In the typical fluctuations regime, $H \sim t^{1/3}$, the PDF of the height at large time is well known to
be described by the Tracy-Widom distributions
\bea \label{TW12} 
&& P_{{\rm droplet}}(H,t)  \simeq \frac{1}{t^{1/3}} f_2\left(\frac{H}{t^{1/3}}\right) \;, \\
&& P_{{\rm flat}}(H,t)  \simeq \frac{2^{2/3}}{t^{1/3}} f_1\left(2^{2/3} \frac{H}{t^{1/3}}\right)  \;.
\label{TW123}
\eea 
If we set $H \sim t >0$ in these formula, we should be probing fluctuations much larger than
$t^{1/3}$ on the right side, where we have obtained above large deviation estimates.
Therefore the large argument behavior of (\ref{TW12}), (\ref{TW123}) should match with the
small $z$ behavior of Eq. (\ref{subdom}).
Indeed, the behavior of the TW-PDF as $x \to + \infty$ is well known
\cite{nadalborot,BaikTW} 
\bea
f_\beta(x) \simeq \frac{\Gamma(1+\frac{\beta}{2})}{\pi (4 \beta)^{\beta/2}} \, x^{(2-3 \beta)/4} \, e^{- \frac{2 \beta}{3} x^{3/2}} 
\eea 
where $\beta=1$ and $\beta=2$ correspond respectively to the flat and the droplet initial conditions. 
Substituting the tails in (\ref{TW12}), (\ref{TW123}) we find (with $H=z t$)
\bea \label{TW122} 
&& P_{{\rm droplet}}(H,t)  \simeq e^{- \frac{4}{3} t z^{3/2} - \ln t - \ln (8 \pi) - \ln z } \;, \\
&& P_{{\rm flat}}(H,t)  \simeq  e^{- \frac{4}{3} t z^{3/2} - \frac{1}{2} \ln t - \frac{1}{2} \ln (8 \pi) - \frac{1}{4} \ln z } \;.
\label{TW1223} 
\eea 
In contrast, starting with the large deviation forms given in (\ref{subdom}) 
and using the exact results for $\chi(z)$ from (\ref{chi_droplet_less}) and (\ref{flat_3})
in the two cases we get
\bea \label{right_drop1}
&& P_{{\rm droplet}}(H,t)  \simeq e^{- \frac{4}{3} t z^{3/2} - \ln t - \ln (4 \pi) - \frac{1}{2} \ln z + 
\ln \Gamma(2\sqrt{z}) } \;, \\  \label{right_flat1}
&& P_{{\rm flat}}(H,t)  \simeq  e^{- \frac{4}{3} t z^{3/2} - \frac{1}{2} \ln t - \frac{1}{2} \ln (8 \pi) 
- \frac{1}{4} \ln z + (\ln 4) \sqrt{z}} \;.
\eea 
Clearly these expressions differ from those in (\ref{TW122}), (\ref{TW1223}) for finite $z>0$,
showing that these large deviation results go beyond the asymptotic large time
regime of Tracy-Widom (and more generally of the Airy processes of the
KPZ fixed point) and carry information about finite time solution. However
in the limit of small $z$, using $\Gamma(2 \sqrt{z}) \simeq 1/(2 \sqrt{z})$, 
we find that they perfectly match as they should.

\section{Left large deviation tail}

We start from the exact relation
\be \label{gumb} 
\langle \exp(- e^{H(t) - t^{1/3} s })\rangle
 =  \langle {\rm Prob}(H < s t^{1/3} - \gamma) \rangle_\gamma 
\ee
where $\gamma$ is a random variable distributed via the Gumbel 
PDF $p(\gamma)=e^{-\gamma  - e^{-\gamma}}$. Therefore
the r.h.s. of (\ref{gumb}) reads
\bea \label{gumb2} 
\int_{-\infty}^{\infty} d\gamma  ~ {\rm Prob}(H < s t^{1/3} - \gamma)  ~ e^{-\gamma  - e^{-\gamma}} \;.
\eea 
On the left large deviation tail the PDF has the form
$P(H,t) \sim e^{- t^2 \Phi_-(H/t)}$ and its associated 
CDF has the same behavior to leading order for large $t$.
Substituting this form in the integral (\ref{gumb2}) leads to
\bea
\int_{-\infty}^\infty d\gamma ~ e^{- t^2 \Phi_-\big(\frac{s}{t^{2/3}} - \frac{\gamma}{t}\big) -\gamma  - e^{-\gamma}}  \;.
\eea 
For large $t$ with $s/t^{2/3}=y$ fixed, one can neglect the $\gamma/t$ term in the
argument of $\Phi_-(z)$, and hence to leading order for large $t$ this
integral is given by $\sim e^{- t^2 \Phi_{-}\left(\frac{s}{t^{2/3}}\right)}$ as discussed
in the main text before Eq. (\ref{painleve1}).

\end{widetext}


\begin{thebibliography}{}
\bibitem{footnote_Fredholm}
We recall that, for a trace-class operator $K(x,y)$ such that ${\rm Tr} K = \int dx K(x,x)$ is well defined, 
$\det(I - K) = \exp{[-\sum_{n=1}^\infty{{\rm Tr \,} K^n}/{n}]}$, where ${\rm Tr}\, K^n = \int dx_1 \cdots \int dx_n K(x_1,x_2) K(x_2,x_3)\cdots K(x_n,x_1)$. The effect of the projector $P_s$ in (\ref{GF}) is simply to restrict the integrals over $x_i$'s to the interval $[s,+\infty )$.
\bibitem{footnote2}
In equation (\ref{F0}) the change of behavior at $y=1/4$ can be simply 
understood as follows. The expansion of the generating function in the left hand side of (\ref{GF}) 
gives $\simeq 1- e^{-t^{1/3} s} \langle e^H \rangle = 1- e^{-t^{1/3} (y-1/12)}$
where we used that $\langle e^{H} \rangle = e^{t/12}$ and $y=s t^{-2/3}$.
Taking a derivative with respect to $s$ then gives the second line of (\ref{F0}).
\end{thebibliography}
\end{document}